\documentclass[12pt]{iopart}
\usepackage[utf8]{inputenc}
\usepackage{iopams}
\usepackage{graphicx}
\usepackage{subcaption} 
\usepackage{float}
\usepackage{tabularx}

\DeclareUnicodeCharacter{200B}{}
\DeclareUnicodeCharacter{2212}{-}
\usepackage[french]{cleveref}

\begin{document}
\title[]
{Thermodynamic Properties and Superstatistics of Graphene under a Constant Magnetic Field}

\author{Yarou M. Assimiou$^1$,  Daniel S. Takou$^2$, Boukari Amidou$^1$, Guingarey Issoufou$^3$, Finagnon A. Dossa$^4$, Gabriel Y. H. Avossevou$^1$}

\address{$^1$Institut de Math\'ematiques et de Sciences  Physiques (IMSP),
	Universit\'e d'Abomey-Calavi (UAC),
	01 BP 613 Porto-Novo, Rep du B\'enin\\
	$^2$Ecole Polytechnique d'Abomey Calavi (EPAC-UAC),
	Universit\'e d'Abomey-Calavi (UAC), Republic of Benin\\
	$^3$D\'epartement de Sciences et Application, Universit\'e de DOSSO, Republic of Niger\\ 
	$^4$Universit\'e Nationale des Sciences, Technologies Ing\'enerie et Math\'ematiques
	(UNSTIM), B.P. 2282 Goho Abomey, Republic of Benin
	}

\ead{ assimiouyaroumora@gmail.com; sabitakoudaniel11@gmail.com; dossafanselme@gmail.com;  amidouboukari12@gmail.com; guingissif@yahoo.fr; gabavossevou@gmail.com}

 \begin{abstract}
 	\par In this paper, we present the solutions of the Dirac-Weyl equation for graphene under a constant magnetic field. The resulting spectrum is used to determine the partition function, a key quantity in the study of thermodynamic properties. From this function, we analyze the mean energy, specific heat, entropy, and free energy in two different frameworks: the canonical ensemble and the superstatistical approach. The study confirms the relativistic nature of electron transport in graphene under a magnetic field. It also reveals that fluctuations introduce additional disorder in the system. The obtained results are in good agreement with those already reported in the literature.
 \end{abstract}

\section{Introduction}

Graphene, a two-dimensional layer of graphite, is a material with remarkable electronic properties that has attracted tremendous interest in both fundamental physics and technological applications \cite{D1'}. Owing to its unique features, such as high electron mobility, thermal conductivity, and mechanical strength, it is considered a promising material for the development of next-generation electronic devices \cite{D1', D2', D3'}.

However, studying electron confinement in graphene presents specific challenges. Due to the Klein paradox, Dirac electrons in graphene cannot be confined by conventional electrostatic potentials. For this reason, magnetic confinement is often considered as an alternative \cite{D4', D5', D6', D7', D8', D9', D10'}.

Several studies have investigated exact solutions of the Dirac-Weyl equation to model the behavior of electrons under various magnetic field profiles. Notable examples include exact solutions for Dirac electrons in exponentially decaying magnetic fields \cite{D7'}, and analytical solutions in the case of constant magnetic fields, hyperbolic wells or barriers, singular trigonometric or hyperbolic fields, as well as complex magnetic fields \cite{D11', D12', D13'}.

Other studies have addressed the exact solution of the Dirac-Weyl equation in graphene under the combined influence of electric and magnetic fields \cite{D25'}, or have analyzed solutions in Cartesian magnetic field configurations \cite{D24'}. The behavior of electrons on a hyperbolic graphene surface under a magnetic field has also been recently examined \cite{Da24'}.

The energy spectrum obtained from the considered quantum equation allows the construction of the partition function, a central parameter in the study of thermodynamic and superstatistical properties of a quantum system. This partition function is then used to determine key thermodynamic quantities such as mean energy, specific heat, entropy, and free energy. These quantities are essential for understanding thermodynamic and superstatistical behavior in quantum physics.

In recent years, increasing attention has been devoted to the analysis of thermodynamic and superstatistical properties due to their broad applications in condensed matter physics, nuclear and molecular physics, and nanotechnology \cite{As7, As8, As9}.

Motivated by the growing interest in graphene and its thermodynamic and superstatistical features, this paper investigates the behavior of graphene under a constant magnetic field, within both the canonical (thermodynamic) and superstatistical frameworks.

This paper is organized as follows: Section~\ref{sec2} presents the method used. The application to the case of a constant magnetic field is developed in Section~\ref{sec3}. Section~\ref{sec4} deals with the analysis of thermodynamic properties in the canonical and superstatistical approaches. Section~\ref{sec5} is devoted to the discussion of the results. Finally, our conclusions are provided in the last section.

\section{Description of the method}\label{sec2} 
\subsection{The Nikiforov-Uvarov (NU) Method}\label{sec1b}

The Nikiforov-Uvarov (NU) Method is a powerful approach for reducing second-order differential equations, through a suitable transformation $x = x(s)$, into a hypergeometric-type equation \cite{D57, D14'}:
\begin{eqnarray}
\Psi''(s) + \frac{\tilde{\tau}(s)}{\sigma(s)} \Psi'(s) + \frac{\tilde{\sigma}(s)}{\sigma^2(s)} \Psi(s) = 0, \label{B1}
\end{eqnarray}
where $\sigma(s)$ and $\tilde{\sigma}(s)$ are polynomials of at most second degree, while $\tilde{\tau}(s)$ is a polynomial of at most first degree \cite{D57, D55, D56}.

The first step of the NU method consists of performing a change of variable using the factorization:
\begin{eqnarray}
\Psi(s) = \phi(s)\varphi(s), \label{B2}
\end{eqnarray}
which transforms Eq.~(\ref{B1}) into \cite{D57}:
\begin{eqnarray}
\sigma(s) \varphi''(s) + \tau(s) \varphi'(s) + \lambda \varphi(s) = 0, \quad \mbox{where} \label{B3}
\end{eqnarray}
\begin{eqnarray}
\sigma(s) = \pi(s) \frac{d}{ds} (\ln \phi(s)), \quad \frac{\phi'(s)}{\phi(s)} = \frac{\pi(s)}{\sigma(s)}, \label{B4}
\end{eqnarray}
\begin{eqnarray}
\tau(s) = \tilde{\tau}(s) + 2\pi(s), \label{B5}
\end{eqnarray}
where $\phi(s)$ is related to a logarithmic derivative, and $\pi(s)$ is a polynomial of at most first degree.
\\
To obtain physically acceptable solutions, \( \tau(s) \) must satisfy the condition:

\begin{eqnarray}
\tau'(s)<0. \label{B6}
\end{eqnarray}
Moreover
 \cite{D57} ,
\begin{eqnarray}
\lambda_{n}=-n\tau'(s)-\frac{n(n-1)\sigma''(s)}{s}, \quad \label{B7} n=0,1,2,\dots
\end{eqnarray}
and a constant \( k \) is introduced such that:

\begin{eqnarray}
\lambda = k + \pi'(s). \label{B8}
\end{eqnarray}
According to the Frobenius method, one also obtains \cite{D57}:
\begin{eqnarray}
\lambda=\lambda_{n}=-n\tau'(s)-\frac{n(n-1)\sigma''(s)}{s}, \label{B9}
\end{eqnarray}where \( \lambda_n \) are the eigenvalues of equation (\ref{B3}).\\
Equations (\ref{B8}) and (\ref{B9}) thus allow the determination of the new energy eigenvalues.\\
Solving the quadratic equation for \( \pi(s) \), using equation (\ref{B8}), yields:

\begin{eqnarray}
\pi(s)=\left(\frac{\sigma'(s)-\tilde{\tau}(s)}{2}\right)\pm\sqrt{\left(\frac{\sigma'(s)-\tilde{\tau}(s)}{2}\right)^{2}-\tilde{\sigma}(s)+k\sigma(s)}, \label{B10}
\end{eqnarray}
\( \pi(s) \) must be a polynomial of degree 1, which is satisfied if and only if:  
\begin{eqnarray}
\Delta = 0, \label{B11}
\end{eqnarray}
that is, the discriminant of the above expression must be zero \cite{D57}.\\
Determining \( k \) is a key step in computing \( \pi(s) \).
\par $\bullet$ \textbf{Determination of the wave function.}\\
We consider equation (\ref{B4}), and according to the Rodrigues formula, we have \cite{D57}:
\begin{eqnarray}
\varphi_{n}(s)=\frac{C_{n}d^{n}}{\rho(s)ds^{n}}\left[\sigma^{n}(s)\rho(s)\right],\label{B12}
\end{eqnarray}
where \( C_n \) is a normalization function, and the weight function \( \rho(s) \) must satisfy the condition:
\begin{eqnarray}
\frac{d}{ds}\left[\sigma(s)\rho(s)\right]=\tau(s)\rho(s).\label{B13}
\end{eqnarray}
\subsection{Reduction of the Dirac-Weyl equation to the Nikiforov-Uvarov equation
}\label{sec1c}
The Hamiltonian of a Dirac electron in Cartesian coordinates is written as follows \cite{D14'}:
\begin{eqnarray}
H = v_F (\sigma \cdot p), \label{1g}
\end{eqnarray}
where \( v_F = \frac{c}{300} \), with \( c \) being the speed of light, \( \sigma = (\sigma_x, \sigma_y) \) are the Pauli matrices, and \( p \) is the momentum operator.\\
Using the minimal coupling rule, that is by replacing \( p \) by \( p + \frac{e}{c} A \), the Dirac equation becomes \cite{D11',D12'}:
\begin{eqnarray}
v_{F}\left(\sigma .p\right)\psi\left(x, y\right)=E\psi\left(x, y\right)\;
=v_{F}\left[\sigma\left(p+\frac{e}{c}A\right)\right]\psi\left(x, y\right). \label{2g}
\end{eqnarray}
Assuming a complex magnetic field \( B(x) \) perpendicular to the graphene surface is applied, we have \cite{D11',D12'}:
\begin{eqnarray} 
A=\left(A_{x}, A_{y}, 0\right) \quad \mbox{and} \quad B(x)=\nabla \times A \label{3g}.
\end{eqnarray}
In the Landau gauge, the vector potential can be chosen as \cite{D15',D16'}:
\begin{eqnarray}
A(x)=\left(0, A_{y}(0), 0\right) \quad \mbox{and} \quad B(x)=\left(0, 0, B_{x}\right)=\frac{dA_{y}}{dx}=A'_{y}(x), \label{4g} 
\end{eqnarray} with $ B(x)\in \mathbb{C} $.\\ In this gauge, the problem can be solved exactly.\\
The superpotential is given by \cite{D11', D12'}:
\begin{eqnarray}
W(x)=k_{1}+\frac{e}{c\hbar}A(x), \quad \mbox{with} \quad W^{'}(x)=\frac{e}{c\hbar}\frac{dA_{y}}{dx}=\frac{e}{c\hbar}B(x).\label{5g} \end{eqnarray}
The associated non-Hermitian Hamiltonians are \cite{D11', D12'}:
\begin{eqnarray}\label{6g}
H^{\pm}=-\frac{d^{2}}{dx^{2}}+V^{\pm}(x),\quad
\mbox{with} \quad V^{\pm}=W^{2}\pm W^{'}(x) ,
\end{eqnarray} which are the complex potentials arising from supersymmetry (SUSY).\\
The corresponding Schrödinger equation is given by \cite{D17'}:
\begin{eqnarray}
\frac{d^{2}\Psi(x)}{dx^{2}}+\frac{2m}{\hbar^{2}}\left[E_{\pm}-V_{\pm}\right]\Psi(x)=0\label{8g}.
\end{eqnarray}
The eigenvalues of the Dirac-Weyl equation (\ref{2g}) are given by \cite{D11', D12'}:
\begin{eqnarray} E_{n}=\pm \hbar v_{F}\sqrt{E_{-}},\label{9g}
\end{eqnarray}with \( n \) a positive integer \((n \in \mathbb{N})\).\\
Positive sign energies correspond to electrons, while negative sign energies correspond to holes \cite{D12'}.\\
The Nikiforov-Uvarov (NU) Method, or its parametric version, will allow us to easily solve equation (\ref{8g}) (if the conditions are met) in order to determine the eigenvalues of the Dirac-Weyl equation (\ref{9g}).\\
The two-component spinor wave function is given by \cite{Da24'}:
\begin{eqnarray}
\Psi_{n}(x,y) = \exp(ik_{1}y) \left(
\begin{array}{c}
\Psi_{n}(x) \\
i \Psi_{n+1}(x)
\end{array}
\right).
\end{eqnarray}

 \section{Application with a Constant Magnetic Field}\label{sec3}
 From equation (\ref{4g}), to obtain a constant magnetic field perpendicular to the graphene plane along the \( z \)-axis, we choose \( B_{z}(x) = (0,0,B_{0}) \). The corresponding vector potential is then \( A_{y} = B_{0} x \). This well-known case, corresponding to the Landau levels, is discussed in several references \cite{D2',D3'}.\\
 The superpotential is given by:
 \begin{eqnarray}
 W(x)=k_{1}+\frac{eB_{0}}{c\hbar}x=k_{1}+Dx \quad \mbox{and} \quad  W'(x)=D=\frac{eB_{0}}{c\hbar}. \label{a1g} 
 \end{eqnarray}  From equation (\ref{6g}), we get:
 \begin{eqnarray}
 V_{-}(x,D)&=&\left(k_{1}+Dx\right)^{2}-D=k_{1}^{2}+D^{2}x^{2}+2k_{1}Dx-D,  \label{a2g}
 \\
 V_{+}(x,D)&=&\left(k_{1}+Dx\right)^{2}+D=k_{1}^{2}+D^{2}x^{2}+2k_{1}Dx+D. \label{a3g}
 \end{eqnarray}For \( 2m = \hbar = 1 \), the corresponding Schrödinger equation is:
 \begin{eqnarray} \frac{d^{2}\Psi(x)}{dx^{2}}+\left[\bar{E}-D^{2}x^{2}-2k_{1}Dx\right]\Psi(x)=0\label{a7g} \;
 \mbox{with}\; \bar{E}=E_{-}-k_{1}^{2}+D.
 \end{eqnarray}
Let \( x = s \Rightarrow dx = ds \), and equation (\ref{a7g}) becomes:
 \begin{eqnarray}
 \Psi"(s)+\left(-\beta^{2}-\gamma^{2}s^{2}-\alpha^{2}s\right)\Psi(s)=0\label{a8},
 \end{eqnarray}
 \begin{eqnarray}
 \mbox{with} \,\; \beta^{2}=\bar{E},\,\; \gamma^{2}=D^{2},\,\; \mbox{and} \,\; \alpha^{2}=2k_{1}D.
 \end{eqnarray}
 By identifying with equation (\ref{B1}), we obtain:
 \begin{eqnarray}
 \tilde{\tau}(s)=0,\,\;
 \sigma(s)=1 \quad \mbox{and} \quad \tilde{\sigma}(s)=-\left(\beta^{2}+\gamma^{2}s^{2}+\alpha^{2}s\right).\label{a9}  
 \end{eqnarray}From equation (\ref{B10}), we have:
  \begin{eqnarray}
 \pi(s)=\pm\left(\gamma s+\frac{\alpha^{2}}{2\gamma}\right) \quad \mbox{for} \quad k=-\beta^{2}+\frac{\alpha^{4}}{4\gamma^{2}}, \label{A12}
 \end{eqnarray} 
 with the condition
  $ \beta^{2}<\frac{\alpha^{4}}{4\gamma^{2}}. $\\
 According to equation (\ref{B5}), we have:
 \begin{eqnarray}
 \tau(s)=\pm\left(2\gamma s+\frac{\alpha^{2}}{\gamma}\right)=
 -2\gamma s-\frac{\alpha^{2}}{\gamma},\label{a13}
 \end{eqnarray}which is the physically acceptable solution, in accordance with (\ref{B6}).\\
 We then have:
 \begin{eqnarray}
 \tau'(s)=-2\gamma<0.
 \end{eqnarray}
 According to equation (\ref{B7}), we have:
 \begin{eqnarray}
 \lambda_{n}=2n\gamma.\label{a15}
 \end{eqnarray}
 According to equation (\ref{B8}), we have:
 \begin{eqnarray}
 \lambda=-\beta^{2}+\frac{\alpha^{4}}{4\gamma^{2}}-\gamma. \label{a16}
 \end{eqnarray}And according to equation (\ref{B9}), we have:
 \begin{eqnarray}
 \lambda=\lambda_{n} \quad \mbox{then
 } \quad 2n\gamma=-\beta^{2}+\frac{\alpha^{4}}{4\gamma^{2}}-\gamma. \label{a17}
 \end{eqnarray}
 After solving, we have:
 \begin{eqnarray}
 E_{-}&=&\epsilon_{n}^{-}=2nD>0 \quad \mbox{and} \quad \epsilon_{0}^{-}=0,\label{a18}
 \end{eqnarray} for all positive integers \( n \).  
 \par \(\bullet\) The eigenvalues of the Dirac-Weyl equation (\ref{2g}) are given by:
 \begin{eqnarray}
 E_{n}=\pm\hbar v_{F}\sqrt{2nD}=\pm\hbar v_{F}\sqrt{n\frac{2eB_{0}}{c\hbar}}.\label{a19}
 \end{eqnarray}The spectrum therefore does not depend on the wave number \( k_{1} \), but only on the magnetic field \( B_{0} \).  
 \par \(\bullet\) \textbf{Determination of the wave functions.} \\  
 According to equation (\ref{B4}), we have:
 \begin{eqnarray}
 \frac{\phi^{'}(s)}{\phi(s)}=-\gamma s-\frac{\alpha^{2}}{2\gamma} \quad \mbox{then} \quad \phi(s)=e^{-\frac{1}{2}\left(D s^{2}+2k_{1}s\right)}.
 \end{eqnarray}
 According to equation (\ref{B13}), we have:
 \begin{eqnarray}
 \frac{\rho'(s)}{\rho(s)}=-2\gamma s-\frac{\alpha^{2}}{\gamma} \quad \mbox{then} \quad \rho(s)=e^{-\left(D s^{2}+2k_{1}s\right)}.
 \end{eqnarray}They represent Hermite polynomials, and according to (\ref{B12}), we have:
 \begin{eqnarray}
 \varphi_{n}&=&\frac{C_{n}d^{n}}{\rho(s)ds^{n}}\left[1^{n}\rho(s)\right] \simeq H_{n}(z) \quad \mbox{with}\quad
 z=\sqrt{\frac{1}{2}\left(D s^{2}+2k_{1}s\right)}
 \end{eqnarray}
Finally, according to equation (\ref{B2}), the final form of the wave function can be written in terms of Hermite polynomials:  
 \begin{eqnarray}
 \Psi_{n}(s)=C_{n}e^{-z^{2}}H_{n}(z).
 \end{eqnarray}

\section{Thermodynamic Properties and Superstatistics}\label{sec4}

\subsection{Thermodynamic Properties}

Several thermodynamic properties can be studied starting from the partition function defined as:
\begin{eqnarray}
Z(\beta) = \sum_{n=0}^{\lambda} e^{-\beta E_{nl}}, \label{e1}
\end{eqnarray}
where $\lambda$ is the highest value of the vibrational quantum number obtained from the numerical solution \cite{D81, Ass}:
 $\frac{E_{nl}}{dn}=0, \; \beta=\frac{1}{kT}$ where $k$ and $T$ are the Boltzmann constant and the absolute temperature, respectively.\\
 The summation in (\ref{e1}) can be replaced by an integral in the classical limit \cite{D81, Ass}:
 \begin{eqnarray}
Z(\beta)=\int_{0}^{\lambda}e^{-\beta E_{nl}}\, dn.\label{T59}
\end{eqnarray} 
From equations (\ref{T59}) and (\ref{a19}), we have:
\begin{eqnarray}\label{5.2}
Z(\beta)=\int_{0}^{\lambda}e^{-\beta(\pm\hbar v_{F}\sqrt{2nD})}\, dn.
\end{eqnarray} 
In the case of electrons, we have:
\begin{eqnarray}
Z(\beta)=\int_{0}^{\lambda}e^{-\beta\hbar v_{F}\sqrt{2nD}}\, dn= \frac{e^{-\beta a}(-1-\beta a)+1}{\beta^{2}b},\label{Z}
\end{eqnarray} with $D=\frac{eB_{0}}{c\hbar}, \; a=\hbar v_{F}\sqrt{2D\lambda} \; \mbox{and }\; b=D\hbar^{2}v_{F}^{2}.
$
\begin{enumerate}
	\item Average vibrational energy
	
	The average vibrational energy is given by	
\begin{eqnarray}
U(\beta)=-\frac{\partial \ln(Z(\beta))}{\partial\beta}=\frac{2-e^{-\beta a}\left(a^{2}\beta^{2}+2a\beta+2\right)}{\beta\left(1-e^{-a\beta}-a\beta e^{-a\beta}\right)}.\label{U}
\end{eqnarray}
\item Heat Capacity

The heat capacity is given by:

\begin{eqnarray}
C(\beta)&=&k\beta^{2}\frac{\partial^{2} \ln(Z(\beta))}{\partial\beta^{2}} \\
C(\beta)&=&-k\bigg[-2+e^{-\beta a}\left(a^{3}\beta^{3}-\beta^{2}a^{2}+4a\beta+4\right)\nonumber\\
&+&e^{-2\beta a}\left(-\beta^{2}a^{2}-4a\beta-2\right)\bigg]\bigg/\left(-1+e^{-a\beta}+a\beta e^{-a\beta}\right)^{2}.\label{C}\nonumber
\end{eqnarray}
\item Vibrational Entropy

The vibrational entropy is given by:
\begin{eqnarray}
S(\beta)&=&k\ln(Z(\beta))-k\beta\frac{\partial \ln(Z(\beta))}{\partial\beta},\\
S(\beta)&=&k\left[\ln\left(\frac{1-e^{-\beta a}-\beta ae^{-\beta a}}{\beta^{2}b}\right)+\frac{2-e^{-\beta a}\left(a^{2}\beta^{2}+2a\beta+2\right)}{\left(1-e^{-a\beta}-a\beta e^{-a\beta}\right)}\right]\label{S}\nonumber.
\end{eqnarray}
\item Vibrational Free Energy

The vibrational free energy is given by:
\begin{eqnarray}
F(\beta)=-\frac{1}{\beta}\ln(Z(\beta))=-\frac{1}{\beta}\ln\left(\frac{1-e^{-\beta a}-\beta ae^{-\beta a}}{\beta^{2}b}\right)\nonumber.\label{F}
\end{eqnarray}
 \end{enumerate}

\subsection{Superstatistics Properties }

Superstatistics is a statistical framework developed to describe driven, non-equilibrium systems characterized by fluctuations in intensive parameters, such as the inverse temperature \( \beta \), chemical potential, or energy \cite{sth1, sth2, sth3, sth4}. These fluctuations occur over spatiotemporal scales and are typically captured by extending the conventional Boltzmann factor into a more general form known as the \textit{effective Boltzmann factor} \cite{sth4, sth5, sth6, sth7, Ass2}.

In this context, superstatistics can be viewed as a superposition of different local equilibrium statistics. The standard approach involves taking the Laplace transform of the probability density function \( f(\beta') \), which results in the generalized Boltzmann factor \cite{sth5, sth7, Ass2}:

\begin{equation}
B_E(\beta) = \int_0^\infty e^{-\beta' E} f(\beta', \beta) \, d\beta'.
\end{equation}

When \( f(\beta', \beta) \) is modeled by a Dirac delta function \( \delta(\beta - \beta') \), the integral simplifies, and a deformation parameter \( q \) can be introduced to yield the generalized Boltzmann factor:

\begin{equation}
B_E^{(q)}(\beta) = e^{-\beta E} \left(1 + \frac{q}{2} \beta^2 E^2 \right).\label{B}
\end{equation}

Here, \( q \in [0, 1] \) is a deformation parameter that quantifies the departure from classical Boltzmann-Gibbs statistics. In the limit \( q \rightarrow 0 \), standard statistical mechanics is recovered.

The partition function in the superstatistical framework is then defined as:

\begin{equation}
Z_s = \int_0^\infty B_E^{(q)}(\beta) \, dn\label{sthe1}.
\end{equation}




These generalized functions are valid for all values of \( q \) and explicitly depend on the system’s energy spectrum. Thus, superstatistics provides a robust framework for analyzing complex, non-linear, and far-from-equilibrium systems where traditional statistical mechanics falls short.\\
Using equations (\ref{B}),  (\ref{sthe1}) and Mathematica, the superstatistics partition function equation is given as :
\begin{eqnarray}
Z_s(a,b,\beta,q)=\frac{2+6q}{a^2 \beta^2}\label{sthe2}.\label{Zs}
\end{eqnarray}

Using the Superstatistic  partition function in Eq.(\ref{sthe2}), the  Superstatistics   thermodynamics properties are obtained as follows.

\begin{enumerate}
	\item   Vibrational mean energy 
	
	\begin{eqnarray}
	U_s(a,b,\beta,q)=-\frac{\partial}{\partial \beta}\ln Z_s(a,b,\beta,q)=\frac{2}{\beta}.\label{Us}
    \end{eqnarray}
 
 \item  Free energy  
 \begin{eqnarray}    
 F_s(a,b,\beta,q)=-\frac{1}{\beta}\ln{Z_s(a,b,\beta,q)}=-\frac{1}{\beta}\ln\bigg[\frac{2+6q}{a^2 \beta^2}\bigg].\label{Fs}
  \end{eqnarray}
  
\item Entropy   
\begin{eqnarray}
S_s(a,b,\beta,q)&=&k\ln Z(a,b,\beta,q)-k \beta \frac{\partial(\ln Z(a,b,\beta,q))}{\partial \beta}\\
&=&k\bigg(2+\ln\bigg[\frac{2+6q}{a^2\beta^2}\bigg]\bigg).\label{Ss}
 \end{eqnarray} 
 \item  Specific heat capacity 
 \begin{eqnarray} C_s(a,b,\beta,q)=k\beta^2\frac{\partial^2}{\partial \beta^2}\ln Z_s(a,b,\beta,q)=2k.\label{Cs}
 \end{eqnarray}
 
 \end{enumerate} 
\section{Numerical Results and Discussions}\label{sec5}
We now analyze the numerical results related to the thermodynamic properties and superstatistics associated with the studied model.
 \begin{figure}[H]
 	\centering
 			\includegraphics[width=7cm, height=6cm]{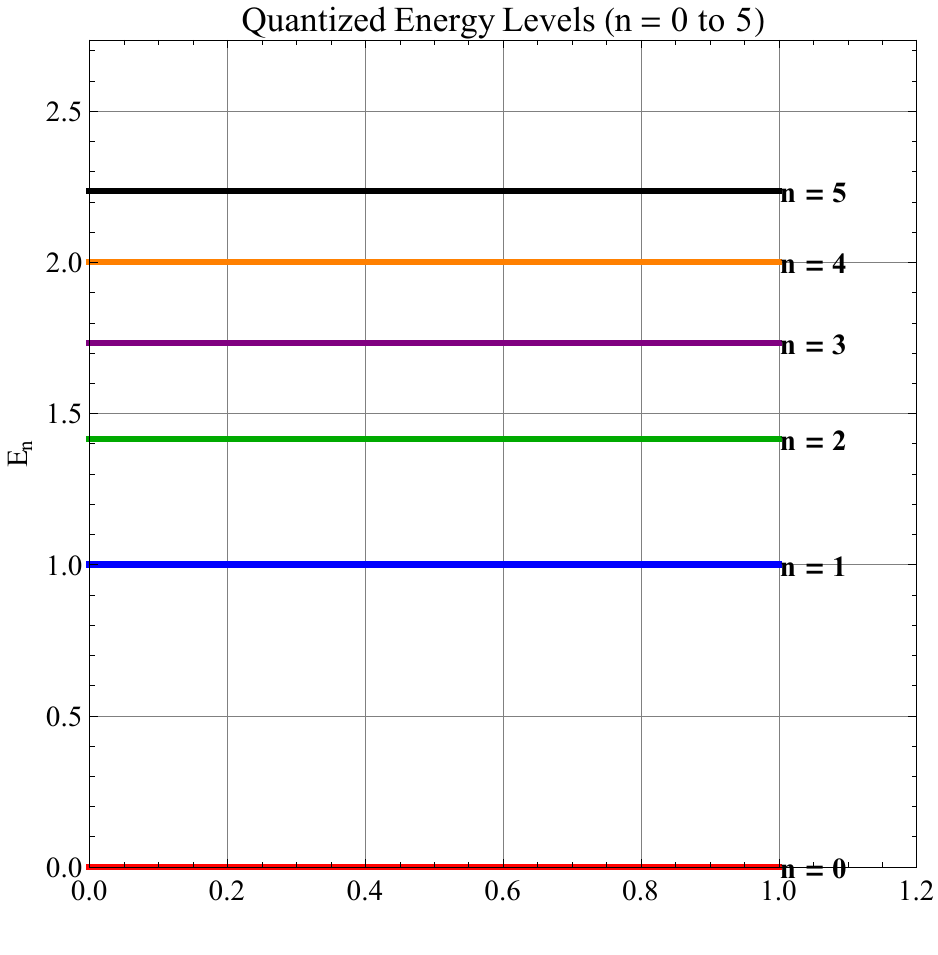}\quad	\includegraphics[width=7cm, height=6cm]{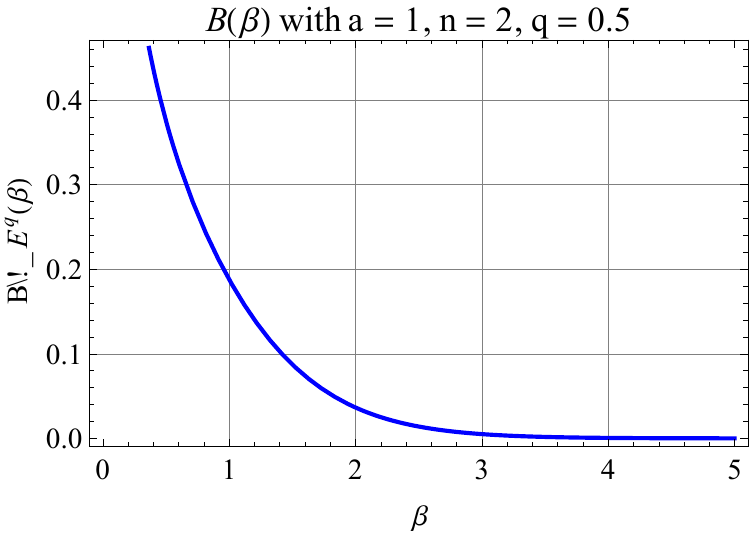}
 	\caption{Variation of the energy spectrum from equation (\ref{a19}) on the left and that of the generalized Boltzmann factor from equation (\ref{B}) on the right.
 	}\label{fig:1}
 \end{figure}

 \begin{figure}[H]	
 	\centering		\includegraphics[width=7cm, height=6cm]{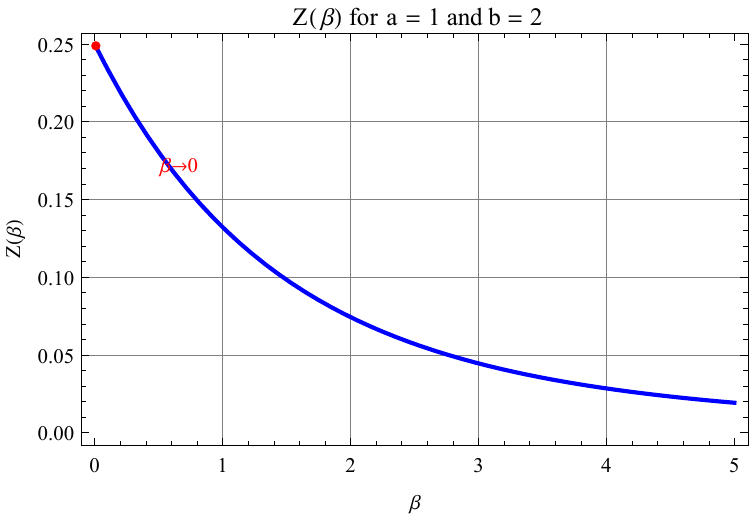}\quad		\includegraphics[width=7cm, height=6cm]{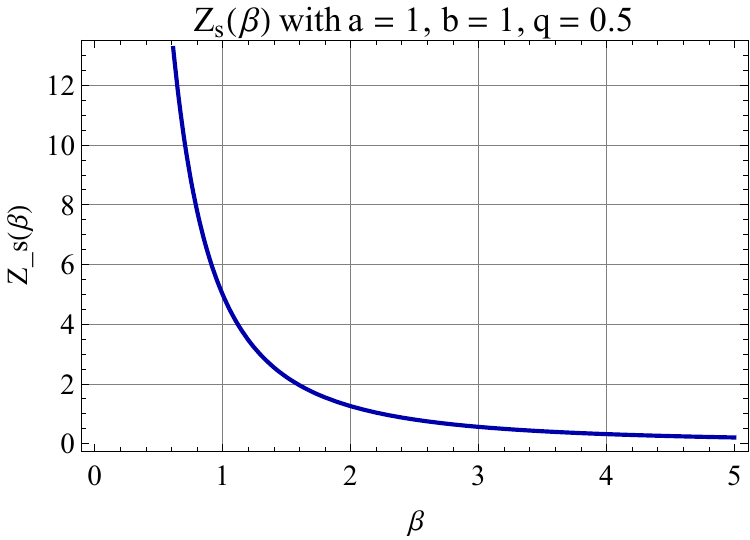}
 	\caption{Variation of the partition function in the classical framework on the left (equation (\ref{Z})) and in the superstatistical framework (equation (\ref{Zs})) on the right.}\label{fig:2}
 \end{figure}

 \begin{figure}[H]	
 	\centering		\includegraphics[width=7cm, height=6cm]{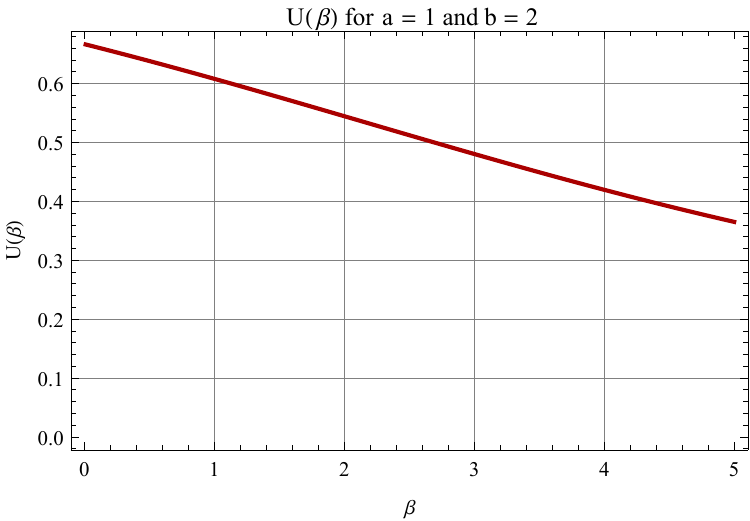}\quad		\includegraphics[width=7cm, height=6cm]{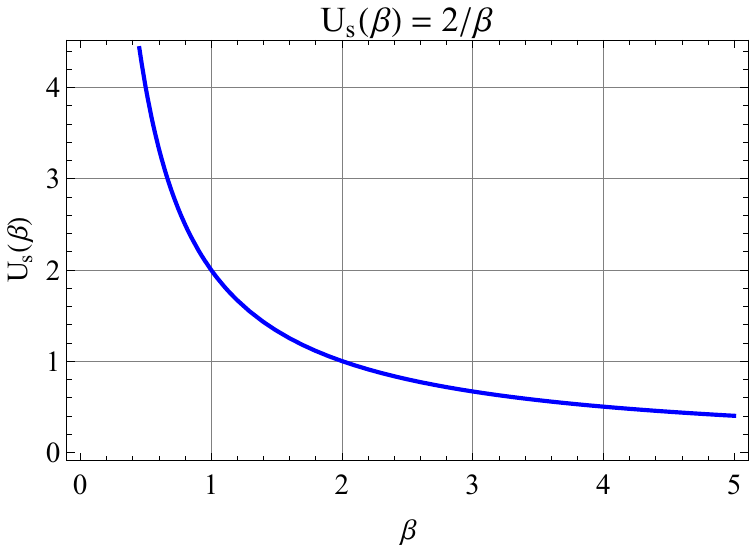}
 	\caption{Variation of the mean energy in the classical framework on the left (equation (\ref{U})) and in the superstatistical framework (equation (\ref{Us})) on the right.
 	}\label{fig:3} \end{figure}

 \begin{figure}[H]	
 	\centering		\includegraphics[width=7cm, height=6cm]{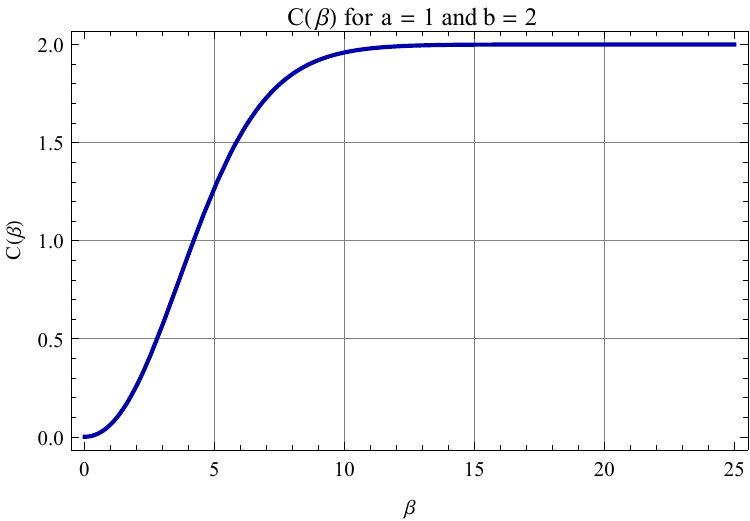}\quad		\includegraphics[width=7cm, height=6cm]{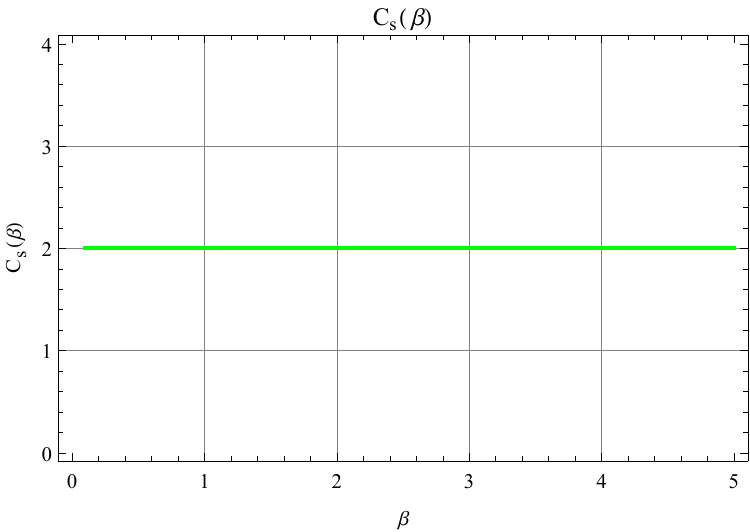}
 	\caption{Variation of the specific heat in the classical regime (equation (\ref{C})) on the left and in the superstatistical regime (equation (\ref{Cs})) on the right.
 	}\label{fig:4}
 \end{figure}

 \begin{figure}[H]	
 	\centering
 			\includegraphics[width=7cm, height=6cm]{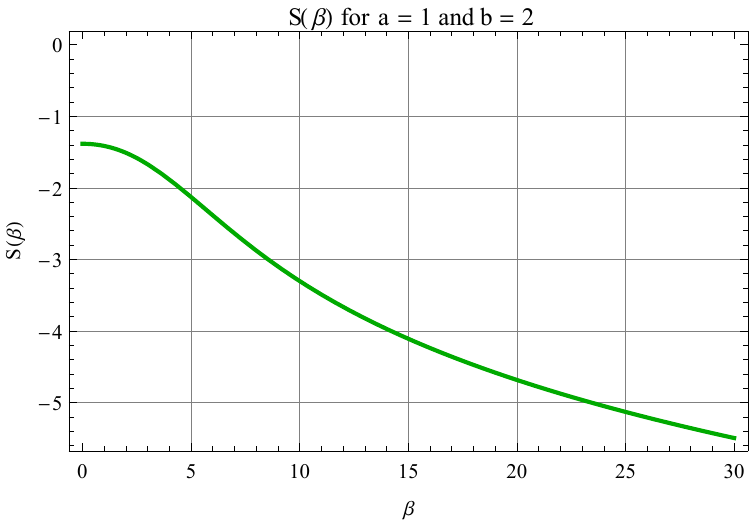}\quad	\includegraphics[width=7cm, height=6cm]{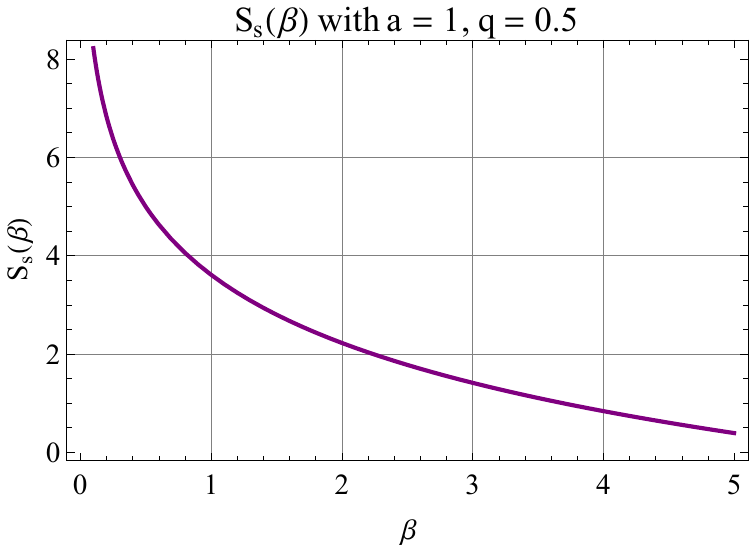}
 	\caption{Behavior of the entropy in the classical regime (equation (\ref{S})) on the left and in the superstatistical regime (equation (\ref{Ss})) on the right.
 	}\label{fig:5}
 \end{figure}

 \begin{figure}[H]	
 	\centering		\includegraphics[width=7cm, height=6cm]{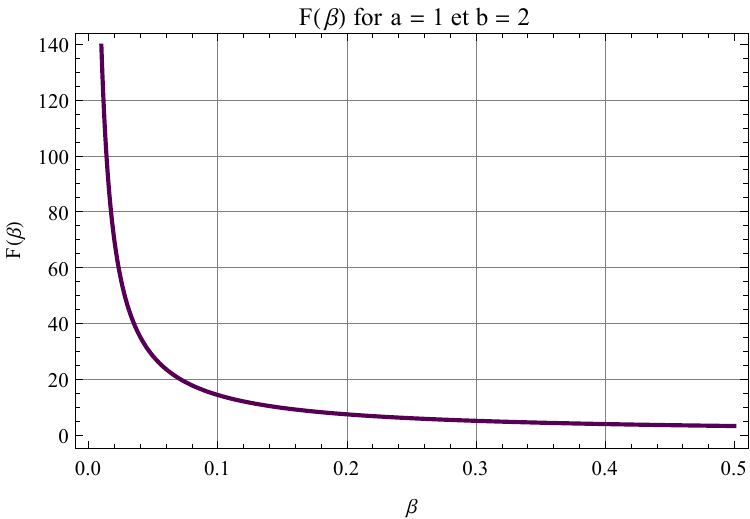}\quad		\includegraphics[width=7cm, height=6cm]{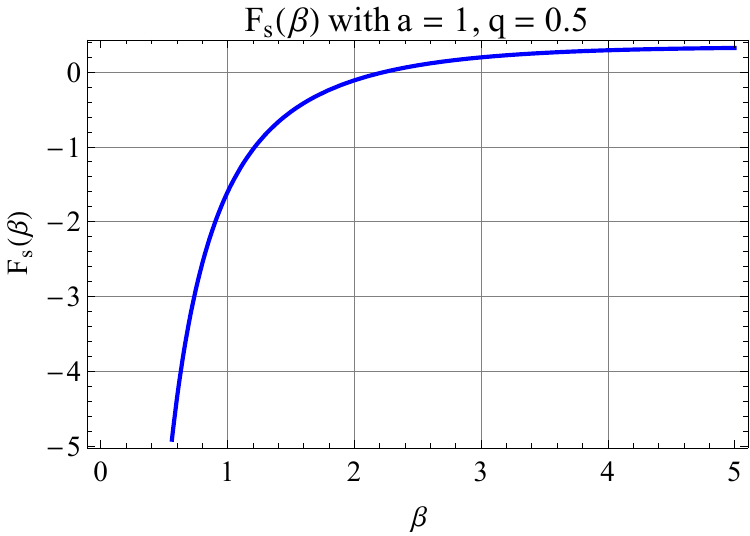}
 	\caption{Variation of the free energy in the canonical framework (equation (\ref{F})) on the left and in the superstatistical framework (equation (\ref{Fs})) on the right.
 	}\label{fig:6}
 \end{figure}
  \par Figure~\ref{fig:1} illustrates the variation of the energy spectrum (on the left) as well as the behavior of the Boltzmann factor (on the right). It is observed that the energy levels are quantized, forming a discrete sequence of discontinuous and non-equidistant values. This indicates that in the presence of a uniform magnetic field, electrons in graphene exhibit relativistic-type quantization. The level $n=0$ is symmetric between electrons and holes, which is a distinctive feature of graphene compared to classical semiconductors. As for the generalized Boltzmann factor, it decreases exponentially with $\beta$: the higher the $\beta$, the lower the probability of occupying an excited level. At high temperature (i.e., for small values of $\beta$), this function tends toward a maximum.
  
  \par Figure~\ref{fig:2} highlights the difference between the partition function $Z(\beta)$, arising from a system with a discrete relativistic spectrum, and its superstatistical counterpart $Z_s(\beta)$, which accounts for thermal fluctuations. While $Z(\beta)$ decreases exponentially, $Z_s(\beta)$ exhibits a slower decay, typical of out-of-equilibrium or non-extensive systems.
  
  \par Figure~\ref{fig:3} shows the variation of the mean energy $U(\beta)$. In the canonical framework (on the left), it decreases gradually with $\beta$. In contrast, in the superstatistical regime (on the right), the mean energy $U_s(\beta)$ decreases more rapidly, following a $\beta^{-1}$ law, which highlights the strong influence of local thermal fluctuations in non-homogeneous systems.
  
 \par In Figure~\ref{fig:4}, a rapid increase in $C(\beta)$ is observed with increasing $\beta$, followed by stabilization around a maximum value, indicating thermal saturation. In the superstatistical framework, $C_s(\beta)$ remains constant for all values of $\beta$, showing independence with respect to local thermal fluctuations.
 
 \par According to Figure~\ref{fig:5}, $S(\beta)$ decreases rapidly with increasing $\beta$, reflecting a reduction in the microscopic disorder of the system as the temperature decreases. $S_s(\beta)$ also decreases. The higher value of $S_s(\beta)$ compared to $S(\beta)$ indicates that fluctuations introduce additional disorder, which is relevant for complex materials such as graphene.
 
 \par According to Figure~\ref{fig:6}, in the canonical framework, a rapid decrease in the free energy $F(\beta)$ is observed as \( \beta \) increases. This behavior reflects the reduction in the energy available to perform useful work as the temperature decreases. The free energy tends toward a value close to zero for large values of \( \beta \), which is consistent with the physics of quantum systems with discrete spectra: at low temperatures, the system predominantly occupies its ground state. 
 
 \par In the superstatistical framework, the free energy \( F_s(\beta) \) exhibits a smoother behavior: it is initially negative for small values of \( \beta \), then gradually increases and tends toward zero as the temperature approaches zero. This behavior highlights the effect of temperature fluctuations accounted for in the superstatistical approach, which smooths out the thermal variations of the system. Indeed, superstatistics introduces an averaging over an ensemble of thermal microstates, which softens the decay of the free energy.
  
  \section{Conclusion}
  We have solved the Dirac-Weyl equation for graphene subjected to a constant magnetic field. The resulting energy spectrum is quantized, forming a discrete and non-continuous sequence. Based on this spectrum, we have studied the thermodynamic properties in both the canonical and superstatistical frameworks.
  
  The superstatistical approach provides greater thermodynamic stability to the graphene system under a constant magnetic field by mitigating the extreme thermal effects observed in the canonical framework. It thus offers a more robust description of thermodynamic properties, particularly in contexts where local thermal fluctuations or out-of-equilibrium situations play a significant role.
  
  The results obtained are in good agreement with those reported in the literature.

   \end{document}